# Title: Softening the "crystal-scaffold" for life's emergence

**Authors:** Gargi Mitra-Delmotte[1*] and Asoke Nath Mitra[2*]

[1]39 Cite de l'Ocean, Montgaillard, St.Denis 97400, REUNION.
e.mail : gargijj@orange.fr

[2]Emeritius Professor, Department of Physics, Delhi University, INDIA; 244 Tagore Park, Delhi 110009, INDIA;
e.mail : ganmitra@nde.vsnl.net.in

*Both are corresponding authors: GM-D for Biology and ANM for Physics related issues.

**Abstract:**
Del Giudice's group study how water can organize on hydrophilic surfaces forming coherent domains (loaning energy from the quantum vacuum), plus quasi-free electrons, whose excitations produce cold vortices, aligning to ambient fields. Their electric and magnetic dipolar modes can couple to oscillatory (electric-organic-dipoles), and/or rotary (magnetic-mineral-dipoles), besides responding to magnetic potentials. Thus, imprinted electromagnetic patterns of catalytic colloids -- c.f. Cairns-Smith's "crystal-scaffold"-- on their structured water partners could have equipped the latter with a selection-basis for 'choosing' their context-based "soft-matter" (de Gennes) replacements. We consider the potential of the scenario of an external control on magnetic colloids forming in the Hadean hydrothermal setting (of Russell and coworkers) -- via a magnetic-rock-field-- conceptually enabling self-assembly, induction of asymmetries, response effects towards close-to-equilibrium dynamics, associative-networks, besides providing a coherent environment for stabilizing associated symmetry-broken quanta, and their feedback-interactions with those of coherent water-domains, to address the emergence of metabolism and replication.

Keywords: magnetic field-control, FeS mineral colloids, soft-matter, water coherent-domains, electromagnetic potential, organic 'takeover', origins-of-life

# 1 Introduction:

**Revisiting Cairns-Smith's 'arch'-metaphor for soft living systems**

Life's efficient use of the colloid-gel state, combining the best from the structured world of solids with the softness and fluidity of the liquid state strongly supports its origins in such a setting [1,2,3,4]. Again as the cell's synthetic machinery reveals a hierarchy of organic compounds of progressively increasingly complexity, approaches to the emergence of life have traditionally centered on organic-based assemblies. These approaches can broadly be distinguished in terms of the targeted location on the machinery: the core comprising simpler organics woven in cycles of small metabolic reactions, or the periphery, consisting of organics with specialized functions. Typically, theories proposing that the metabolic wing of life had preceded the replicating one focus on the simpler core reactions that could have been hosted on assemblies of either organics [5] or colloidal inorganic compounds [6], as running the simple (core) cycles in the reverse (reductive) mode can shed light on the plausible links between the transfer of electrons from higher to lower states of energy and the emergence of life in the Hadean conditions of disequilibrium [7,8,9]. 'Replicators' in 'metabolism-first' approaches, are thought to have gradually emerged as feedback type correlations emerged between networks, eventually leading to the code [10,11]. These scenarios are based on the sound premise that the efficient capture of energy, released from the primordial reaction networks, would have awaited the emergence of order in the hosting-surfaces [12]. Still, plausible stages as to how this came about and enabled the quantitative reproduction of information do need to be chartered. At the other extreme, theories proposing that the replicating wing appeared first focus on the complex organics on the periphery of the reaction nets. Although their principle of targeting the function is a sound one, these proposals face a dilemma on account of the assumed association of these functions with complex organic structures (whose synthesis they attempt) as their peripheral location indicates that they must have arisen later. And, even if it were possible to accomplish the Herculean task of construing a suitable geo-chemical setting in the Hadean that could be argued to meet the requirements for producing ample amounts of the complex building blocks--directly from very basic organic units,--the conceptual bridge as to how the complex biomolecules associated with bio-functions came to be formed from the simpler organics in the core, would remain unaddressed. By uncoupling the organic structure-function association, Cairns-Smith [13] proposes the analogy to the 'arch' metaphor as a way out of this impasse, suggesting that in the Hadean, available non-organic predecessors that could have accomplished these functions may have started the cycle, and thus offered a functional basis of selection for takeover by the bio-organics. His novel theory posits that clay minerals were the first self-replicating living forms where genetic information was encoded in patterns like random sequences of stacked mineral layers (instead of nucleotide bases), crystal growth defects, aperiodic distributions of ions, etc. Its main hurdle is that the rigid framework of hard crystals is seen as conceptually incompatible with evolving dissipative dynamical structures.

Nevertheless, the intuitive argument of a "readily available" mineral scaffold [13] could be expanded in the sense of a lowest common denominator (LCD)-template having the basic functional attributes of living systems in their simplest form, where a synthesis of mineral-based approaches, could explain how energy (associated with metabolism) and information (associated with replication) became synonymous in complex bio-matter [14]. For instance, although crystalline order is typically associated with rigid geometrical patterns, we need to study the carriers of its information, to look for an aspect of its ordering that could be compatible with motion, towards the dynamical interception of energy by catalytic colloidal surfaces. It is therefore pertinent to look at the correspondence between the "global motions" of enzyme components and the collective vibrations of the crystal lattice [15]. Also, the capacity to capture the released energy and stretch its residence time, and prevent its fast dissipation to the sink, is difficult to imagine in terms of far-from-equilibrium examples, such as turbulent flows that have no obvious coherent connection between the microscopic and macroscopic flows. Now, a coherent ordering mechanism associated with an inorganic colloidal scaffold could have compatibility with both requirements, and this extended soft template (c.f. [13]), *with an essential role for structured water* as a fountainhead for life, forms our premise. For meeting this demand, we need a mechanism providing long range correlations as in condensed matter physics (CMP) via association of ordered structures with the colloidal-gel state.

**QFT/QED: A basis for correlated components and interaction space**

A glimpse of a kind of stable yet dynamic order in an open system that could have enabled such interception can be had from the many-body model proposed by Umezawa et al [16, 17] which states [18]: "in any material in CMP any particular information is carried by certain ordered patterns maintained by certain long range correlations mediated by massless quanta. It looked to me (Umezawa) that this is the only way to memorize some information; memory is a particular pattern of order supported by long range correlations." Indeed, Frohlich's [19] landmark proposal suggests how a continuous supply of energy should lead to the establishment of coherent order over macroscopic distances via long-range phase correlations between molecules. Upon pumping, excitations of the densely packed particles with electric and elastic interactions are expected to lead to vibrations building up into collective modes of photons, phonons, etc. this generates a coherent dipolar wave motion which exchanges energy with the surrounding electromagnetic field. And, being coherent, this energy gets stored in an ordered manner (not thermalized). Importantly, the electromagnetic field coupled to the dipolar bio-molecules plays a key role in mediating their coupled interactions.

Now this picture of correlated components connected via an environmental (electromagnetic) field, obviously calls for a paradigm shift from the conventional picture of independent system parts that could be isolated from each other, and also the environment to keep thermal effects at bay. Traditionally complex bio-systems are dealt with either thermodynamic or causal-dynamical approaches. Now while holistic approaches like the former miss out on microscopic details in their search for global principles, reductionist ones like the latter lose sight of holistic features like coherence. Furthermore, features like coherence demand a knowledge of the underlying manifolds of the interactions, i.e. the structure of the space in which the components are interacting. Although causal-dynamical approaches can provide such a description, too much 'reductionism' associated with them interferes with a holistic formulation; in contrast, thermodynamics is way too macroscopic for adequate descriptions of underlying interacting fields. In this context, quantum electrodynamics (QED)-- the acknowledged 'Queen' of all quantum field theories (QFTs)-- has a natural in-built holistic feature without losing out on causality which is needed to address hierarchial description all the way from microscopic to macroscopic domains [20]. (For a flavour of this vast subject the reader is referred to a tutorial [21] elsewhere). As noted by the Del Giudice group, the field-environment, crucial for the dynamics of the correlated particles, finds a powerful representation in the QFT language of 'vacuum' (originally formulated for empty space). Now while a modern form of QFT predicts the existence of infinitely many minima (degenerate vacuum) the former nevertheless owes its origin to its simplest form, namely the so called unique vacuum—the single state of lowest energy. The time evolution of these degenerate vacua on the other hand, follows a more complex pattern of symmetry principles than the one followed by system- components in empty space (governed by Euler-Langrangian equation). And, different patterns of time-evolution of the vacua, characterized by breaking of conventional symmetries -- especially time-reversal -- as compared to that of the system- components in empty space, allow for a description of the evolution of the system across its vacua in an irreversible (bio-like) fashion [22, 23, 24]. In this scenario, the 'phase transitions' of the system are seen as typical QED symmetry-breaking effects corresponding to the different vacua that are imprinted on the dynamics of its components [25].

      Del Giudice and Tedeschi [26] further argue that the archetypal framework of a crowded chemical reactor of randomly colliding independent molecules (range of few Angstroms only), cannot explain how, despite crowding, living matter components detect each other from afar and interact via long sequences of biochemical reactions, with sharply defined space-time order. Also, in a chemical reactor released energy gets dissipated as heat, i.e. increased kinetic energy of randomly moving molecules, leading to increase in entropy and wild temperature fluctuations. In contrast, in living matter, where energy gets stored for carrying out different functions, its supply seems to fit in with a wave-like QED scenario (see Sect.2) of a moving electromagnetic field whose frequency depends on the coherence strength of the medium, and therefore can attract `co-resonating' molecules. In this way, as energy gets captured in a coherent medium, its coherent domain expands (by adding more and more field-coupled molecules), while fresh electromagnetic oscillations are added to the existing ones; thus by mutual feedback the process multiplies rapidly. This scenario ensures growth and order at very low entropy [27] at almost constant temperature (Sect.2). And by virtue of the fact that the intensity of the coupled electromagnetic field is inversely proportional to the difference of their oscillation frequencies, so that only molecules whose oscillation frequencies are nearly resonant attract themselves strongly, a dynamic basis to the origin of the selective-recognition codes orchestrating the bio-reactions (hitherto accepted as dogma), is achieved [28].

**Outline of paper**

These insights lead to the intriguing possibility that the colloidal state can also show properties associated with crystal lattices, and when seen from this angle the two approaches seem to converge, with colloidal minerals participating in a functional template as a basis for selection of bio-molecules—from simpler building units to increasingly complex ones (like membrane phospholipids, proteins, carbohydrates, nucleic acids, etc)—and also their mutual interactions. Indeed, this is in addition to their capacity to act as catalytic soft surfaces, compatible with dynamics. Thus the following may be outlined as:

1) We briefly review the scenario of water domains organized on hydrophilic colloidal surfaces envisaged by Del Giudice and coworkers [29] as to how these dissipative structures forming by coherent dynamics, can show a reducing nature; and effects electromagnetic potentials (instead of fields) can exert on a system with long-range phase correlations, before concluding this section with a brief historical perspective on QED developments (**Sect.2**)

2) Water being the common denominator in the hypothesized Hadean scenario and in the present day life forms, we consider how mineral colloids can provide a correspondence with the collective response behaviour [30] of complex biological molecules. Analogous to H-field induced associative networks with ferrofluids, we look at the 'mound scenario' [7] where magnetic iron-sulphur colloids can be similarly controlled with a (spatially varying) field from magnetic rocks [31, 32]. In a thermodynamic description, the symmetry-broken patterns of hypothesized field-driven mineral colloid assemblies and ATP-driven soft bio-matter are compared (**Sect.3**)

3) We point out how the presence of two different dipolar modes could have implications for water CDs to have acted as the "scaffold" enabling 'takeover' (**Sect.4**)

4) We consider how the paradigm of a soft "scaffold" (c.f. [13]) of field-controlled catalytic-mineral-colloids enrobed in organized water [26] could address the entangled emergence of both (metabolic and replicating) wings of life (**Sect.5**)

5) Conclusions and scope (**Sect.6**)

**2. Dissipative coherent domains and vector potential effects**

Having appreciated the relevance of a coherent system of phase-correlated components, for being in a position to intercept energy as in life processes, one also sees the huge jump needed for a molecular assembly (as in conventional origin-of-life approaches) — an uncorrelated ensemble — to make, in order to achieve that status. To that end, Del Giudice and coworkers [29, 27, 28, 25] consider the unique role of water in living systems (that accounts for 70% in weight and 99% in molar weight). In fact, the ordering of water, leading to what are called exclusion zones in the presence of structure makers like PEG or natural ones like carbohydrates, has for long been exploited for the stabilization of bio-molecules for *in vitro* and *in vivo* purposes [33]. In contrast to ordinary water, interfacial water close to hydrophilic bio-surfaces, studied in detail by Pollack's group, has been found to have a number of anomalous properties. Apart from resisting penetration of solutes, this exclusion zone (EZ) is associated with charge separation; the structured zone itself is negatively charged while protons concentrated in the region beyond are free to diffuse, as dictated by the local electric gradient [3]. The boundary between EZ and normal water forms a redox pile thanks to the negative electric potential of the former of the order of 100 mV relative to the latter. Remarkably, the growth of this zone has been found to be induced upon exposure to light radiation, thus making this energetic pathway a central protagonist for the origin of life. What's more, fluorescing exclusion zones show the possibility of electronic transitions, with implications for proto-metabolic processes; recall that energy of the order of soft-X-rays is required to ionize a water molecule (ionization potential 12.60 eV [29]. And the diffusing protons -- a by-product of the thus built-up exclusion zone-- enable the process of coalescence by balancing the inter-particle repulsion (owing to negatively charged structured water envelopes on the bio-molecules), forming ordered arrays. This is indeed consistent with the observation that increasing the light intensity diminishes the distance between micro-particles [3]. But, this scenario of orderly motion is at variance with the conventional picture of water with uncorrelated rotations and thermal effects [34], which is typically described with two-body interactions at relatively large distances. On the other hand, at much smaller relative distances, the opposite scenario dominates. Namely, the N-body interactions become increasingly dominant as r becomes smaller, when radiative corrections acquire increasing dominance giving rise to time-dependent e.m. fluctuations (necessarily coherent as against thermal effects).

These observations therefore have a natural interpretation in terms of the QED-based 'coherence domain' (CD) theory [26, 35] in which charged system components are coupled to the e.m. field. CD theory makes use of a new collective ground state of the structured water (with entrapped e.m.fields) whose features differ from those of the isolated molecular ground state water (the incoherent gas phase). The non-coherent and coherent states are separated by an energy gap ($\lambda_{CD}$ = hc / $E_{ex}$ = 0.1m), corresponding to the infrared region, where $E_{ex}$ = 12.06 eV is the energy of the excited state of water molecules). In this manner, this ubiquitous molecule gives rise to a 'coherence trap', eventually leading to a larger coherence unit [26, 34], where the water CD offers a reservoir of almost free electrons that are excitable at each step of metabolism with a concomitant reduction of entropy. Now QFT, as distinguished from mere (single-particle) quantum mechanics, is characterized by an uncertainty relation connecting the number N of field quanta (unspecified) and a `phase' $\Phi$ which-- unlike the usual phase of wave-like motion in quantum mechanics-- describes something more subtle, namely the (coherent) `rhythm' (*a la* Del Giudice) of the field oscillations as a whole. This uncertainty relation [36] in QFT which reads as:

$$\Delta N \Delta \Phi \geq \hbar/2,$$

plays an important part in determining the collective coherence measure ($\Phi$) of the aggregate of the bio-particles (vis a vis their number N) under the influence of the ambient e.m. field. A second uncertainty relation brings in the zero-point energy, or the energy of the quantum vacuum, which is best illustrated by the harmonic oscillator problem in quantum mechanics, viz., $\hbar\omega/2$, plays a subtle but crucial role in energy book-keeping in the physics of bio-systems. This concept shows up rather dramatically when a large number of (incoherent) particles undergo a `phase transition' down to an ordered state with a decrease in entropy. This process must be accompanied by the release of a certain amount of heat energy (so as not to violate the Second Law of thermodynamics). And the source of this energy is precisely the zero point quantum vacuum energy noted above.

As another interesting observation, Ho et al [37] have inferred the presence of coherent domains in embryos serving as detectors for the Aharanov-Bohm effect in a suitably designed set-up. The abnormality profiles of these embryos match those exposed to weak static magnetic fields, pointing to the sensitivity of the former batch to the *vector potential* in an essentially field-free region. The significance of this subtle effect has been interpreted by Brizhik et al [38] in terms of *non-linearity* of the corresponding Schroedinger equation. For a quick derivation, reinterpret the usual momentum p as *kinetic momentum*, on the lines of its energy counterpart which admits of a division in terms of kinetic and potential energies. Now the corresponding `potential momentum' involves the e.m. *vector potential A* in the form eA where e is the charge of the concerned particle so that the kinetic momentum becomes (p – eA), and the usual kinetic energy term $p^2/2m$ term gets replaced by $(p-eA)^2/2m$. Now, for a linear (perturbative or $1^{st}$ order) dependence on A, the phase $\phi$ can be simply 'gauged away' (i.e., trivially eliminated), so that only a non-coherent description of the state remains! But with a more complete (non-linear) dependence on A, one has the more interesting possibility that the phase $\phi$ can no longer be eliminated and depends directly on A, thus producing a coherence effect, while the `mundane' amplitude $\psi_0$ has a more or less `classical' (non-coherent like) dependence on the fields E and H [38]. Note also that the phase $\phi$ is directly connected to the Aharanov-Bohm effect, (recall the definition of its characteristic phase shift) !

To sum up the findings with a historical perspective, it is good to recall that the unique vacuum was the source of fundamental discoveries in the middle of the last Century, such as the `Lamb shift', which had generated novel ideas of Renormalization of the Vacuum, as well as second and higher order electromagnetic corrections, leading to incredible agreement with experiment up to one in $10^{12}$ ! In the excitement of these dramatic discoveries however, the potential powers of the e.m. field for impacting other (less exotic) phenomena somehow got lost sight of ( due to lack of immediate motivation ?). In this respect an important phenomenon—first recognized by the Italian group of Del Guidice and coworkers--concerns the role of QED as a (ubiquitous) *ambient field* making its impact on biomatter at the more earthly `first-order level' itself, even without going into such fancy second order corrections! What Del Giudice's group recognized was that in the ambient interaction of the e.m. field with bio-matter, its vacuum is already in a highly degenerate form, together with its features of symmetry breaking and the presence of long-range Nambu-Goldstone (NG) bosons [39, 27, 28]. This interaction is best expressed in terms of the e.m. potentials A, $\phi$ (which have longer ranges than the corresponding field quantities E and H). It seems surprising that this `elementary form of e.m. interaction with bio-matter which is otherwise quite basic, was not considered earlier in the literature in such a `holistic' form.

## 3. Water organized on bio-matter vs field-controlled mineral colloids

The above framework for stable dissipative structure formation and storage of energy in coherent form brings about a synthesis of energy and information (!) that seems highly relevant for water, which is capable of changing its supra-molecular organization depending upon its interaction with the environment. This is thanks to the special properties of the electronic spectrum of the ubiquitous molecule-- their proposed candidate providing the 'hardware'. As discussed above [38], the vector potential A--the 'brain' behind the coherent arrays (of e.m. field-entrapped CDs)-- already offers a `switch' for controlling the phase of the coherent system, with a range of implications for living organisms, such as global selection mechanisms and communication networks. But, as water in the coherent organized state requires a hydrophilic surface for its structuring, this may well also hold the key to how the 'takeover' from minerals by organics [13] had come came about, its' collaborating presence being a common factor in both scenarios—hypothesized mineral colloids in the Hadean, and organic living systems today. In particular, we refer to the dynamical basis of selection via the e.m.f. mediated resonance-basis of attraction of molecules ([26, 27, 28, 24], Sect.2). Now, since Life is a historical phenomenon, the question of how the phase of the coherent water system could have been manipulated, must hence be intimately entangled with the properties, the nature, the workings of these 'eccentric' bio-molecules that 'took over' from their inorganic ancestors. In addition, non-covalent interactions between the complex molecules, like lipids, proteins, carbohydrates, and nucleic acids, via a variety of recognition modes—that underlie biological language-- frequently appear in myriad conserved patterns for propagating information observable across kingdoms of life. Instead of life's emergence based on the single origins of only one of these complex forms in a suitable geochemical setting (where the issue of how these associations might have happened would have to be either postponed or skipped), the picture gels with that of gradually building up organic networks within a special microenvironment that could allow time for associations between molecules and later between types of molecular networks to occur [10, 11]. And, a dynamic super-scaffold comprising organized water domains (see Sect.2) could have selected different kinds of specialized organics to replace its components carrying out different functional roles; this scenario entails that associations between complex bio-molecules came about by these substitutions.

**3.1 Soft-matter; large response functions: a 'thermodynamic' description**

We start with a phenomenological (*a la* statistical mechanics) description urging that the attributes of what has come to be known as 'soft matter' coined by Pierre-Gilles de Gennes [30] be reviewed in the light of the "principle of biological continuity" (see, e.g., [40]). One could ask if mineral-like clusters (iron sulphur, etc) have been around for 'more reasons' than the huge list already compiled hitherto [41] (see Sect.3.2 next). Had a soft mineral-containing scaffold given residence to a quasi-particle/s, which was equally comfortable in association with its organic successor, providing the same pattern of dynamic functional 'goods'? Again, the huge advances in the materials sciences sector show couplings between different d.o.f.s (magnetic, elastic, thermal, etc), as in those of biomolecular systems [42], and raise the possibility of similar patterns in mineral colloidal particles [43]. In fact, thermally stable mineral liquid crystalline phases are of interest as they can be electron-rich in contrast to organic ones, and therefore possess pronounced electrical, optical and magnetic properties [44]. Now, the collective nature of the "large response functions" [30] underlies the susceptibility of complex bio- matter to small external perturbations. In fact, the main aspects of soft matter: response functions, non-covalent weak interactions and entropic forces play a key role in biological organization and close-to-equilibrium dynamics. Its emergence seems hard to imagine from very simple organics at the core (that could be traced to primordial abiogenics) unless this transition can be in principle addressed in a continuous deterministic manner. Guided by Cairns-Smith's paradigm that any material with this potential was fine so long as it was readily available, we suggest that a template comprising mineral colloids [32] could have realized these functions. In particular, one may recall that *fields* can carry signatures such as sensitive material responses, indicating how a coherent source can supply energy to matter to sustain an isothermal, symmetry-broken aperiodic assembly subject to random fluctuations. At a macroscopic level, a good example was provided by Breivik [45] who showed how magnetic information transfer (via self-assembled 3 mm-sized magnets) helps encode a sequence of independent and identically-distributed random variables [46], underlying life's information-rich aperiodic order [47]. For an extension of this to the nano-scale consider field-induced aggregates observed in ferrofluid dispersions [48, 49], described as a phase-separation of a particle-concentrated phase from a dilute one [50]. These close-to-equilibrium structures (requiring about tens of milli Tesla fields for their formation) are dissipative in nature, breaking up when the field is switched off, and offer a basis for slowly changing patterns as in soft bio-matter. The external H-field breaks the rotational symmetry of the dispersed and disoriented single-domain particles that are subject to thermal fluctuations from the bath, and imposes a directional order. Again, Dyson's [51] use of field-accreted-matter for simulating 'analog-life', gives an instructive edge to magnetism-based proposals, e.g. implementation of Boltzmann machine type of neural network based on inter-spin exchange interactions in a spin glass [52]; and associative memory simulations, using particle-particle dipolar interactions in nano-colloids [53].

**3.2 Rock magnetism and magnetic colloids in 'mound scenario'**

Similarly, as magnetic rocks are a good provider of moderate H-fields for accretion of nanoparticles forming on the Hadean Ocean Floor [31], this brings us to the alkaline seepage site mound scenario [6, 54, 55], illustrated in Figure 1 (reproduced with permission from Russell and Martin [7]; Russell et al [54]). Here, negatively-charged colloidal mineral greigite forming under alkaline mound conditions (as pH well above 3 [56]), does resemble an aqueous-based ferrofluid. Significantly, the key to stabilizing its colloidal-gel state lies with organics [57]. And, the strong structural similarities between iron-sulphur clusters in enzymes and their mineral counterparts that were likely to have been present on the primordial ocean floor, underlies the interest in sea-floor hydrothermal systems [7]. For, clusters of iron-sulphur are seen ubiquitously across living systems (despite their pre-dominantly organic basis), and carry out a variety of roles, such as electron transfer, radical generation, sulfur donation, control of protein conformational changes associated with signal transduction, to name some [58, 59, 60, 61]. In fact, spin polarization and spin coupling are key characteristics of the sulphur bridged complexes, embedded within these ancient bio-constructs [62]. The spins on metal sites in di- to polynuclear iron sulphur clusters are coupled via what is called Heisenberg exchange coupling, which typically favors antiferromagnetic alignment of neighbour-spins. This in combination with valence delocalization within the cluster helps bring about coupling of different degrees of freedom and add to the complexity of the system [62]. Ligand binding can thus affect not only the electronic distribution, but also the pattern of spin alignment, and hence the net total spin state [63]. The complex web of interactions-- orbital interactions, electron delocalization and spin coupling-- in the iron-sulphur clusters [64] shows the likelihood of a similar profile for magnetic mineral colloids in the mound membranes.

In the mound scenario, colloidal membrane surfaces, envisaged as hosting simple metabolic cycles, comprise of iron-sulphide minerals, such as mackinawite and greigite, which along with some other transition metal compounds could have been geo-chemically available for carrying out similar catalytic type of functions, and acted as proto-metabolically relevant catalysts [7, 65, 55]. Not only that, but the confrontation of geo-fluids at different pH across precipitating colloidal FeS membranes, also provides insights into the essential dependence of biological processes on a proton-motive force (chemiosmotic gradient) across cell membranes, which is a more or less invariant mechanism of energy conversion across living forms [66].

The 'mound scenario' thus weaves together the key material patterns highly relevant to life processes—soft colloid gel state, catalytic enzyme-like iron-sulphur clusters (for primordial metabolism), plausible abiogenic reactions (with passage of electrons to lower energy states), natural gradients (of redox, pH, temperature), to name some— into a persuasive geochemical location on the Hadean Ocean Floor. It gives a plausible account of how key abiogenics could have accumulated [67] and coordinated with each other [10] in membranous inorganic compartments, as well as dynamically ordered framboidal reaction sacs [68]; the latter forming through interplay of attractive and repulsive forces (c.f. [69]). Indeed, spherical, ordered aggregates of framboidal pyrite about 5μm in diameter were found in fossil hydrothermal chimneys [70, 71, 72, 73]; see Figure 2 (a) and (b), that seemed to have grown inorganically from the spherical shells of FeS gel. Furthermore, Russell and coworkers [70] have noted the size similarities of the magnetosome crystals to that of pyrite crystallites (~ 100nm in diameter) comprising the interior of framboids that appeared to have grown inorganically from spherical shells of iron-sulphide gel; their earlier stages would have comprised reduced forms of framboidal iron sulphide, just as in iron sulphide bearing bacteria. Here, the quasiperiodic organization associated with naturally formed framboids [74] is intriguing in view of their resemblance to the well-studied dynamic order in laboratory made quasi-crystals. These form via an accretion-based mechanism [75] and show unusual combinations of properties (resembling living matter). Thus, the observation of scale-free framboids rekindles the long-standing interest in these forms for the origins-of-life [76], providing as they do, an opportunity to study the surface-minimized packing of colloidal spheres as independent dynamical components [46]. Wilkin and Barnes [56] explain the formation/stability of micro-meter sized framboids, using an interplay of negatively charged repulsive and magnetically attractive forces, where a size > 100nm would orient crystals to the weak geo-magnetic field ~ 70 microTesla. Extrapolation of this to the milli-tesla scale (via a local field from magnetic rocks) shows the feasibility of *nano-scale aggregates*, analogous to ferrofluid ones. This conjecture [31], inspired by chimneys showing framboids (plate 2 of [73]), gets support from observations of fractal greigite framboids [77] which independently affirm the natural propensity of this mineral towards an aperiodic, nested organization.

As to the possible relevance of these observations to the coherent dynamics discussed in the previous sections, we note in passing that a correspondence exists between fractal structures and the algebra of "q-deformed" coherent states (Fock-Bargmann Representation); see Vitiello [78, 79] for details. While this is beyond the scope of the present paper, this is an interesting way to describe the emergence of long range correlations at the global (mesoscopic/macro) level from local (microscopic) deformation patterns.

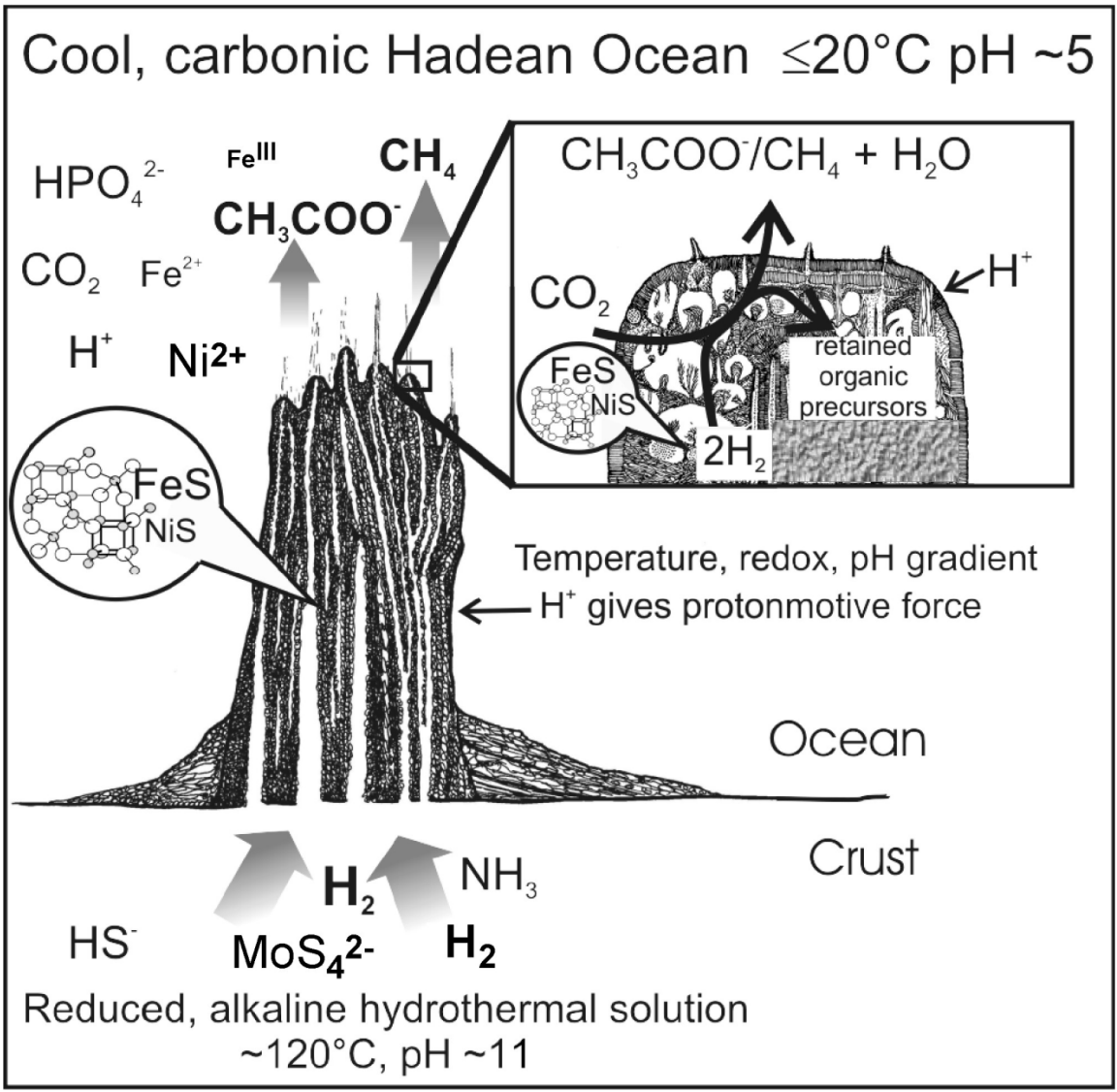

Figure 1. The hydrothermal mound as an acetate and methane generator.

Steep physicochemical gradients are focused at the margin of the mound (see text, also [54] for details). The inset (cross section of the surface) illustrates the sites where anionic organic molecules are produced, constrained, react, and automatically organize to emerge as proto-life (from Russell and Martin [7], and Russell et al [54], with permission).

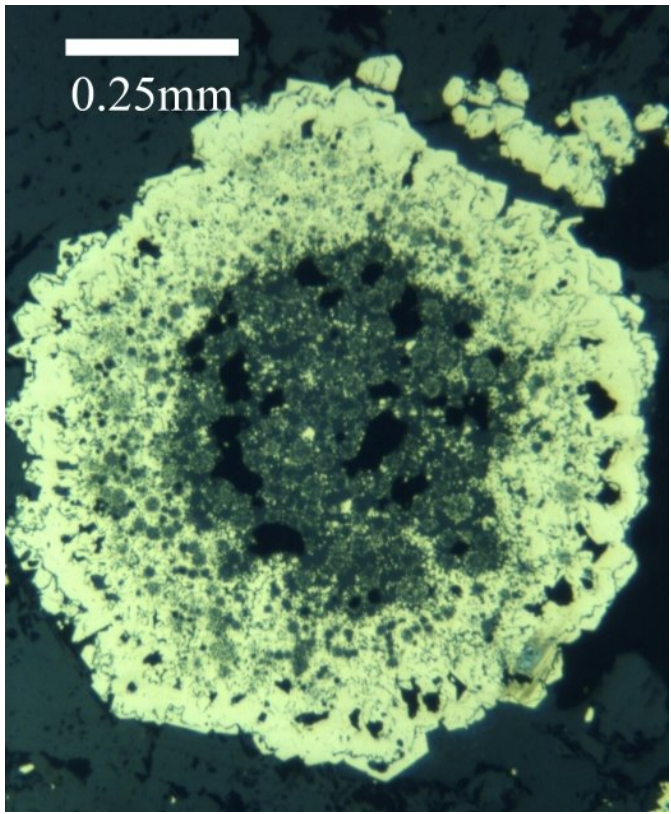 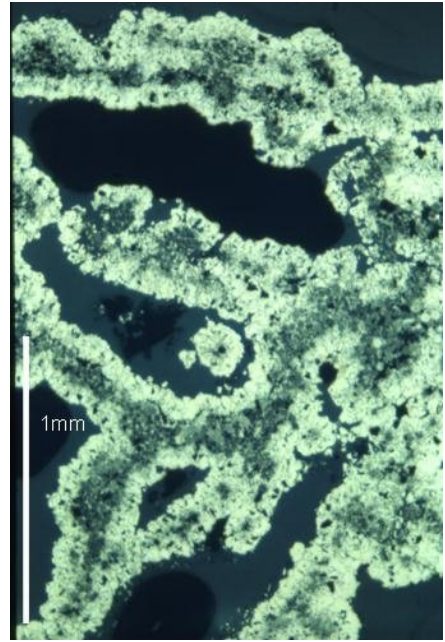

(a) Small vent          (b) Sheaves

Figure 2. Framboids in chimneys:

(a) Small pyrite vent structure: Reflected ore microscopy of transverse section shows a central area of empty black spaces plus (grey) fine framboidal pyrite, and a fine euhedral authigenic rim surrounded by baryte, with minor pyrite; (b) Sheaf system, formed from coalescing rods of anastamosing microcrystalline pyrite. Black areas are empty spaces; central regions are framboidal pyrite with an exterior of crystalline pyrite. (Labelled pictures given by Dr. Adrian Boyce are reproduced with his kind permission; Source: Boyce et al. [71, 72]; Boyce 1990: Exhalation, sedimentation and sulphur isotope geochemistry of the Silvermines Zn + Pb + Ba deposits, County Tipperary, Ireland; Boyce, Unpublished Ph.D. thesis, University of Strathclyde, Glasgow).

### 3.3 Environment as a field—'Taming the noisy Shrew'

Now, the scenario of a magnetic template for 'soft-bio-matter', unfolds a *novel role* for the *environment*. Here, far from being solely a harbinger of perturbative influences, the environment holds the *controls* for the assembly process itself: the field enables the super-paramagnetic particles to overcome thermal fluctuations so that in the resulting aggregates, the components are held together by reversible weak dipolar interactions. Thus, in contrast to crystal lattices, the soft assemblies seem conceptually compatible with dynamics, and would have remained intact, even if the 'layers' of the aggregates were used as surfaces for adsorption (e.g. by abiogenics) or as passages to diffusively migrating particles. For, an H-field can connect the response-to-field variations (i.e. susceptibility) and the correlators between interacting spins via the fluctuation-dissipation theorem (FDT). And, for gradual flux changes to be consistent with effectively isotropic local surroundings for particles (diffusing through the field-induced aggregates), moderate gradients in the mound may have sufficed (c.f. [55]). Sure enough, the weak magnetic dipolar interactions between the latter and those forming the 'layers' have a striking resemblance to the recognition-based movement in transcription and translational processes. Note that the same patterns of diffusive migration through assemblies also map with the ATP-driven migration of bio-molecular motors on cytoskeletal protein networks. Next, to consider coupling of scalar chemical reactions with vector processes like diffusion [80], let's extend surface-transfer reactions envisaged in ligand shells of adjacent transition metal elements on crystal surfaces [81] to 'template'-layers of magnetically ordered particles (c.f. [10]). Now, within a common FDT framework for *asymmetric* movements, binding to non-magnetic ligands (e.g. organics) would increase the net potential energy barrier of the particles for interacting with their 'template'-partners compared to their unbound counterparts. Thus greater diffusive exploration of the organic bound particles (as in bio-molecular motors) contrasts with the entrapment of the unshielded ones into an expanding network of dipolar interactions (as in growth phenomena). Note that the isothermal, cyclic switching between different entropic states in molecular motors, is pictured here using a gentle flux gradient (non-homogeneous rock H-field) periodically perturbed by local H-fields of 'template'-partners, leading to alternating high and low-'template'-affinity states due to the dipole's magnetic degree of freedom [31]. In contrast to these first-order-like phase transitions, the superimposition of a gentle temperature gradient on field-induced structures results in a decrease in magnetization (a second order phase transition) of a diffusing dipole (c.f. [82]). Here, changes in susceptibility seem a plausible response mechanism for harnessing various fluxes in the mound, enveloped by further colloidal/mineral precipitates. In this context, the coupled d.o.f.s in bio-matter ([42], see Sect.3.1) have a strong resemblance to the secondary effects of magnetism in a substance arising as a consequence of couplings between its different physical properties : magneto-caloric, magneto-electric, magneto-optic, magneto-striction [83]. Hence similarly, via the fluctuation-dissipation response mechanism, these couplings could have enabled the environment to harness any gentle gradients using its' field as a "switch". In addition, the passage of charges is expected owing to gradients in the mound, which would induce a magnetic field. The scenario of the latter interacting with the rock field at the base of the mound, and leading to a (reversible) homopolar-motor-like rotation of conducting assembly components, does seem reminiscent of the rotatory ATP-driven pumps that are thought to be amongst the early-evolved mechanisms.

### 3.3.1 Chirality : Naaman; Rosenberg; mound scenario

All along, the constructive use of noise (via FDT) is possible, thanks to the magnetic asymmetry of the assembly. Their ATP-driven soft-matter counterparts in living systems, however, employ chiral molecules for achieving ratchet-effects. A causal connection to understand this correspondence could be given in terms of spin-polarized electrons (as in [84]). This is also consistent with the geochemical-based view of life's origin [54], which posits that life emerged on the surface of a wet planet (likely earth) once the early evolved mechanisms were set in place (mechanisms requiring light, such as photosynthesis, therefore evolved later in this scenario, which makes it relevant to consider interactions with chirality ingredients like circularly polarized light for later stages of evolution). Briefly, the work of Naaman and coworkers [84, 85] reveals how co-operative effects can endow assemblies of chiral dipolar molecules with the capacity to selectively discriminate between spins pointing 'up' or 'down'. A similar spin-filtering effect is achieved in spintronics by applying an external field to induce magnetization in ferromagnetic thin films. On the other hand the magnetization in its' biological counterpart—layered organization of dipolar chiral molecules-- is based on two stages: 1) the magnetic field-created by transfer of charge (electron or hole) through chiral molecules aligns the magnetic dipole of the charge transferred; 2) then exchange interactions in the layered domain keeps them aligned (the spin–orbit interaction is negligible here). Furthermore, in an important study, Rosenberg [86] has shown how chirality can affect reactions in the presence of spin-polarized electrons, provided by radiating the magnetic substrate binding the organics. The requirements for producing low energy spin-polarized electrons -- ionizing radiation and a magnetic substrate—seem to match with the 'infrastructure' available at the base of the mound. In addition to the presence of high levels of radioactivity and the reducing conditions prevailing in the Hadean Earth's crust, Ostro and Russell (unpublished results 2008), have described how iron and iron-nickel particles derived from the subhypervelocity flux of iron and NiFe-metal containing chondritic meteorites and micrometeorites, and distributed on the ocean floor, would be partially scattered throughout the mound (see [31]). The invasion of the mound by magnetized extraterrestrial materials would have been further reinforced by secondary magnetization of the magnetite produced as an alteration product at the base of the mound [31] as a consequence of serpentinization [87]. The plausibility of having spin-polarized electrons, in the mound then throws up the intriguing correspondence between the available carriers of the spin-polarized electrons -- the field-ordered magnetic dipolar colloids [31, 32] and structured water [3, 29, 27, 28, 24] —and the assemblies of chiral dipolar biological replacements which show magnetic behaviour upon polarized charge transfer [85]. Indeed, recent results demonstrate that mackinawite, which is isostructural with high-temperature Fe-based superconductors, also shows similar magnetic profiles, thus leading to the conjecture that this important component of the 'mound scenario' may well be one of the simplest among Fe-based superconductors [88].

## 4. QFT scenario: Two modes of water CD-"scaffold"

Albeit, described above (Sect. 3) at a `stat mech' (thermodynamic) level, one repeatedly encounters analogies of collective behaviour (many coupled spins vs many coupled atoms), where symmetry broken patterns in inorganic colloids show matching correspondences with their biological counterparts. This scenario is fully compatible with a QFT description, with merely a change from imaginary (Matsubara) `time' to real time, *a la* Feynman, the system, as the carriers of ordering information. With the `ordering' there is effectively a freezing of the d.o.f.'s of some elementary (microscopic) components, leading to a coherent, collective (macroscopic) quantum behaviour called the ordered pattern, (see [21, 39, 24]). One can also check that the rotational symmetry broken N-G-bosons (magnons) seem to be more relevant to dynamical biological systems (see the application of rotational symmetry-breaking via electric dipoles in the dissipative brain model of Vitiello [89]), rather than the translational symmetry broken phonons-- N-G bosons in the crystal context. Nevertheless, the quasi-particles associated with the symmetry-broken organization patterns could hold the key to how the 'takeover' [15] from inorganic-magnetic dipolar to organic-electric dipolar nature came about, the association with interfacial water CDs being common to both scenarios.

We have already seen in Section 2, how organized water can exhibit crystal-like collective properties, where a deep reshuffling of its electron clouds allows it to act as an electron donor. Water CDs can act like machines and collect energy –even non-coherent sources like thermal noise—and store in a coherent form, and even include "guest molecules" [26]. Further, the association of coherent water CD's with a reservoir of quasi-free electrons enables the production of metastable coherent excited states [90]. Thus, when excited by some externally supplied energy, the ensemble of quasi-free electrons execute rotational motion (as cold vortices) with quantized magnetic moments that can align to ambient magnetic fields; their stability is ensured by their "cold" coherent nature, preventing thermal decay [26]. Thus the water CDs can be associated with both electric (oscillatory) and magnetic (rotatory) dipoles. Now, the criterion for a "guest" molecule to gain entry into the CD "machine" is that of a matching frequency to that of the CD host via its (lowest order) radiative dipole. And, interfacial water CDs—capable of both kinds of absorption/emission exchanges-- can provide a powerful scaffold mechanism for swapping between inorganic (magnetic-rotatory dipolar) and organic (electric-oscillatory dipolar) colloidal organizations. This is since being coherent systems their phase can be tuned by an external e.m. potential [38, 35]. Thus, for consistency with the principle of biological continuity, e.m. potentials associated with quasi-particles, arising from the various symmetry-breaking contexts in biological organization today, had to be similarly linked to mineral-based ancestors in the Hadean, for enabling the envisaged 'takeover' [13]. Again, the correspondences between the hypothetical Hadean and biological scenarios (Section 3) provide functional contexts for energy dissipation; here the thermodynamic-FDT-basis of description can be checked for their compatibility with coherent exchanges with water CDs. The latter can increase their coherent stores even by accepting incoherent thermal noise, the requirement for release being that the "guest" provides a matching frequency (see above). Evidently though, the organics could not 'takeover' all the functions, such as electron transfers and sensing (see below).

## 5. Metabolism, Replication, and Sensing

Magnetic networks also provide a conceptual platform for bringing together a variety of mechanisms—such as harnessing different fluxes (since further colloidal/mineral precipitates could envelop the mound)—as an alternative to say, awaiting the emergence of the elaborate genetic set-up enabling lateral gene-transfers. For in principle the presence of magnetic-dipolar modes in water CDs offer a mechanism for a framboid-like scaling up of magnetic assemblies, permitting their navigation to zones with weaker field intensities. And we speculate that once the weak geo-magnetic field started to develop, similarly organized organic replacements could have disembarked from the mound. The former also offers a plausible response-based mechanism for navigation. Note the similarity between the passage of a magnetosome containing bacterium, moving in response to the extremely-low-frequency geo-magnetic field [91] (although by sensing changes in the inclination magnitude increasing from the equator to the poles) and the *scaled-down scenario* of ligand-bound super-paramagnetic particles traversing a field-induced aggregate in response to (gentle changes in flux lines due to) an external moderate H-field [31]. For, bacteria do use these cell-components as compasses [92, 93, 94] that help to orient them to the geo-magnetic field, as they navigate through waters. Interestingly, magnetosomes are composed of ordered magnetite crystals, as well greigite ones, albeit less frequently [95, 96]. Their ancient origins can be appreciated from the conjecture that "Magnetoreception may well have been among the first sensory systems to evolve" based on the universal presence of single-domain crystals of magnetite across many species and groups of organisms, ranging from bacteria through higher vertebrates that exploit the geo-magnetic field for orientation, navigation, and homing [97, 98]. In addition, different magnetite-based models of biological magnetic-torque transducers have been proposed as a basis for sensing and transducing magnetic-field changes via direct/indirect coupling to mechanosensitive ionchannels [99]. Indeed, to quote from Kopp and Kirschvink [100], "The magnetosome battery hypothesis suggests the possibility that the magnetosome arose first as an iron sulfide-based energy storage mechanism that was later exapted for magnetotaxis and still later adapted for the use of magnetite".

The gradual substitutions of colloidal magnetic networks bound by weak dipolar interactions could conceptually explain how recognition-based communication language (e.g. complementary base-pairing, enzyme-substrate, etc) may have arisen in water-coated bio-networks bound by non-covalent, weak recurring bonds (H-bonds, van der Waals); note the capacity of the latter to undergo sol-gel transformations, like reversible magnetic gels. Further, as structure (shape, polarity, etc) dictates interactions, one wonders if incorporation of substitutes had not been an information registering mechanism of a coherent ancestor leading to the structure-function associations across biology, the 'contents' being different functions in different structural (memory) addresses. Now the emergence of a digital form of storing information in this scenario—where associative networks were available [31]—is likely to have been a later development that can be understood in terms of connections developing between networks [10, 11]. Nevertheless, a magnetic environment proposed by us (see [101]) can provide a basis [32] for Patel's [102] proposal of a quantum search leading to a digital mechanism, if a coherent ancestor indeed brought forth life, that is. Other workers have also provided arguments involving iron-sulphur minerals [103] and the colloid-state [104] that support a quantum-basis for the origins of life.

## 6. Conclusions and Scope

Cairns-Smith's crystal scaffold is softened via field-responsive colloids to conceptually access the 'Replicator' in two stages: 1) An environmental field enabled the assembly, induction of asymmetries, response effects for close-to-equilibrium dynamics, associative networks of magnetic colloids, besides providing a coherent environment for stabilizing associated (a variety of) symmetry broken quanta and their feedback-interactions with those of the coherent water-domains. 2) Analogue processes in a coherent water-mineral-colloid assembly are speculated to have enabled the 'leap' to digital form of information processing. In this new scenario, the catalytic activity and soft-matter patterns arose from field-responsive mineral colloids, whereas energy capture (from proto-metabolic reactions, and a variety of fluxes) and storage in far-from-equilibrium dissipative structures was throughout associated with the coherent water domains organized on the colloidal inorganic or organic surfaces. This might provide insights into the entangled origins of the replicating and metabolic wings of life.

Indeed, in this manner, Nobel Laureate Albert Szent-Gyorgyi's view of life being "an electron looking for a place to rest" can be seen to be consistent with $CO_2$ as its eventual destination [55] and the essentially organic nature of life. Again, in the words of the father of modern biochemistry, if "water was dancing to the tune of solids" --with tunes as the associated e.m. potentials-- this "dance" would have continued uninterrupted if organic replacements of the inorganic ancestors had played the same "tune".

**Acknowledgements:** We are grateful to Prof. Michael Russell and Kirt Robinson for bringing the work of the Naaman's group to our notice; and Dr. Adrian Boyce for kindly providing his labeled framboid pictures. We are grateful to the Referee for his careful reading of the manuscript, leading to precise insertions of appropriate references. We thank Prof. Michael Russell for inspiration and support (data and figures, plus key references, e.g. Brizhik et al 2009; Carmeli et al 2002; Kwon et al 2011). We are grateful to Dr. Jean-Jacques Delmotte for providing financial and infrastructural support.